Are papers addressing certain diseases perceived where these diseases are prevalent? The proposal to use Twitter data as social-spatial sensors

Lutz Bornmann*, Robin Haunschild**, & Vanash M. Patel***

*Division for Science and Innovation Studies
Administrative Headquarters of the Max Planck Society
Hofgartenstr. 8,
80539 Munich, Germany.
Email: bornmann@gv.mpg.de

** Max Planck Institute for Solid State Research
Heisenbergstraße 1,
70569 Stuttgart, Germany.
Email: r.haunschild@fkf.mpg.de

*** Department of Surgery and Cancer
10th floor, Queen Elizabeth the Queen Mother Wing
St. Mary's Hospital, London
W2 1NY
and
Department of Colorectal Surgery
West Hertfordshire NHS Trust
Watford General Hospital
Vicarage Road, Watford
Hertfordshire, WD18 0HB
Email: vanash.patel06@imperial.ac.uk

**Abstract**

We propose to use Twitter data as social-spatial sensors. This study deals with the question whether research papers on certain diseases are perceived by people in regions (worldwide) that are especially concerned by the diseases. Since (some) Twitter data contain location information, it is possible to spatially map the activity of Twitter users referring to certain papers (e.g., dealing with tuberculosis). The resulting maps reveal whether heavy activity on Twitter is correlated with large numbers of people having certain diseases. In this study, we focus on tuberculosis, human immunodeficiency virus (HIV), and malaria, since the World Health Organization ranks these diseases as the top three causes of death worldwide by a single infectious agent. The results of the social-spatial Twitter maps (and additionally performed regression models) reveal the usefulness of the proposed sensor approach. One receives an impression of how research papers on the diseases have been perceived by people in regions that are especially concerned by the diseases. Our study demonstrates a promising approach for using Twitter data for research evaluation purposes beyond simple counting of tweets.

# 1   Introduction

Citations have been used to measure impact of papers over decades. This exclusive focus on citations in research evaluation has changed in recent years (Bornmann, 2016; Bornmann & Haunschild, 2017). Alternative metrics (altmetrics) have been proposed to measure impact broader, not only on academia itself. The overview of altmetrics definitions by Erdt, Nagarajan, Sin, and Theng (2016) show that there is no formal definition of altmetrics; the definitions vary slightly. The definitions agree, however, that altmetrics have been proposed as an alternative or supplement to traditional bibliometrics and are based on various data sources (e.g., Twitter or Mendeley). With separate conferences (see http://www.altmetricsconference.com) and journals (see https://www.journalofaltmetrics.org), altmetrics seem to emerge as a sub-field in scientometrics "with a broad investigative community focused in the exploration of theoretical, empirical, and procedural aspects" (González-Valiente, Pacheco-Mendoza, & Arencibia-Jorge, 2016, p. 236). According to Moed (2017), there are three drivers of the expansion of the altmetrics field: (1) the policy domain which is interested in impact of research beyond academia, (2) new developments of information and communication technologies which facilitate social interactions, (3) the Open Science movement with the goal to make scientific activities more transparent and better accessible.

Haustein (2016) distinguishes between seven main types of altmetric data sources which focus on specific types of social activity each:

(a) "social networking (e.g., Facebook, ResearchGate)

(b) social bookmarking and reference management (e.g., Mendeley, Zotero)

(c) social data sharing including sharing of datasets, software code, presentations, figures and videos, etc. (e.g., Figshare, Github)

(d) blogging (e.g., ResearchBlogging, Wordpress)

(e) microblogging (e.g., Twitter, Weibo)

(f) wikis (e.g., Wikipedia)

(g) social recommending, rating and reviewing (e.g., Reddit, F1000Prime )" (p. 417).

In this paper, we focus on Twitter which is a source that has been frequently used in altmetrics research hitherto. Twitter is a web-based microblogging system enabling users to post short messages (Mas-Bleda & Thelwall, 2016). It is a decisive advantage of Twitter data that they cannot only be used as tweet counts, but also for social networks (e.g., Haunschild, Leydesdorff, & Bornmann, 2019; Haunschild, Leydesdorff, Bornmann, Hellsten, & Marx, 2019) and spatial maps (Sinnenberg et al., 2017).

The overview by Erdt et al. (2016) reveals that altmetrics research do not only use data from many sources, but also cover a broad spectrum of topics such as the motivation of researchers to use social media, field- and time-normalizations of impact data, visualizations of data, gaming or spamming the data, and distributions of the data across countries, gender or disciplines. Blümel, Gauch, and Beng (2017) identified two main research lines in altmetrics within this broad spectrum: "the first kind of topics are 'coverage studies' of articles with mentions in social media platforms and their intensity … The second type of studies is cross validation studies that employ comparisons of altmetric data sources with traditional measures of scholarly performance such as citations". Based on the observation that already many altmetrics studies have correlated altmetrics and citations, Bornmann (2015) performed a meta-analysis which allows a generalized statement on the correlation between metrics from alternative data sources and citations. His results reveal that "the correlation with traditional citations for micro-blogging counts is negligible (pooled $r = 0.003$), for blog counts it is small (pooled $r = 0.12$) and for bookmark counts from online reference managers, medium to large (CiteULike pooled $r = 0.23$; Mendeley pooled $r = 0.51$)" (p. 1123). Subsequent primary studies have reported similar results.

Most of the studies correlating altmetrics and citations were interested in the question of the meaning of altmetrics: do altmetrics measure something similar or different as citations? Although altmetrics have been already used in the research evaluation practice (see e.g., Colledge, 2014), the meaning of them is not clear. Accordingly, these and similar statements can be frequently found in the scientometrics literature: "Yet at the moment, there is limited understanding of what precisely these indicators mean" (Blümel et al., 2017). "There is no uniform definition, and therefore no consensus on what exactly is measured by altmetrics and what conclusions can be drawn from the results" (Tunger, Clermont, & Meier, 2018, p. 124). Since the meaning of altmetrics is not clear, the terms describing the measured impact varies: societal impact, public attention, non-scholarly popularity, diverse forms of impact, and non-traditional scholarly influence (Konkiel, Madjarevic, & Rees, 2016; Moed, 2017; Waltman & Costas, 2014). Triguero, Fidalgo-Merino, Barros, and Fernández-Zubieta (2018) speak about the "scientific knowledge percolation process" (p. 804), which is defined as the flow from the scientific community to the wider society.

In this paper, we would like to add another term: social-spatial sensor. We use Twitter data to investigate whether research on certain diseases (e.g., tuberculosis) reaches the people that are especially concerned by the diseases (i.e., the regions with many people having the diseases). Since (some) Twitter data contain location information it is possible to spatially map the activity of (some) Twitter users referring to certain papers (e.g., dealing with tuberculosis). The resulting maps reveal whether heavy activity on Twitter is correlated with large numbers of people having certain diseases. Higher correlations can be expected, since – according to Kuehn (2015) – many people "share symptoms or information about health-related behaviors on Twitter long before they ever see a doctor" (p. 2011).

The World Health Organisation (WHO) ranks tuberculosis, human immunodeficiency virus (HIV), and malaria as the top three causes of death worldwide by a single infectious agent. In 2018 tuberculosis caused 1.5 million deaths, HIV caused 770,000 deaths, and malaria caused 405,000 deaths. Although these diseases occur in every part of the world, the largest number of new tuberculosis cases arose in South East Asia and Africa (66%). Similarly, over two thirds of all people living with HIV live in Africa, and this region also carries a disproportionately high share of the global malaria burden (93%). We use these three infections as examples to demonstrate how Twitter data can be used as social-spatial sensors. Identifying relevant articles in the Medline bibliographic database is more accurate when searching single infectious agent diseases as Medical Subject Headings (MESH) rather than diseases caused by multiple pathogens.

## 2  Previous research on Twitter and spatial analyses of online activities

### 2.1  Twitter research

Users on Twitter can use the system to share short messages: "tweets range from one-to-one conversations and chatter, to updates of wider interest about current affairs, encompassing all kinds of information" (Zubiaga, Spina, Martínez, & Fresno, 2014, p. 462). Since tweets can contain (formal or informal) references to scientific papers, the data source might be interesting for research evaluation purposes (Priem & Costello, 2010). Twitter citations are defined "as direct or indirect links from a tweet to a peer-reviewed scholarly article online" (Priem & Costello, 2010). According to Mas-Bleda and Thelwall (2016), Twitter citations of papers reflect "attention, popularity or visibility rather than impact, so that Twitter mentions may be an early indicator of the level of attention (including publicity) that articles attract" (p. 2012).

It seems that topics are frequently discussed on Twitter that are currently trendy such as breaking news or ongoing events (Bik & Goldstein, 2013; Zubiaga et al., 2014). Previous research could not clearly say whether tweets on papers reflect impact mostly from the general public or researchers (Sugimoto, Work, Larivière, & Haustein, 2017; Yu, 2017). For example, Haustein, Larivière, Thelwall, Amyot, and Peters (2014) states that "Twitter is widely used outside of academia and thus seems to be a particularly promising source of evidence of public interest in science" (p. 208). Other studies concluded that mostly researchers tweet about papers: "the majority of tweets stem from stakeholders in academia rather than from members of the general public, which indicates that the majority of tweets to scientific papers are more likely to reflect scholarly communication rather than societal impact" (Haustein, 2019, p. 753). The company Altmetric classifies users tweeting about papers into four groups: researcher, practitioner, science communicator, and member of the public (Yu, 2017).[1] Haustein et al. (2016) define Twitter users with respect to various levels of engagement "differentiating between tweeting only bibliographic information to discussing or commenting on the content of a scientific work" (Haustein et al., 2016, p. 232).

Two topics were in the main focus of Twitter research: (1) correlations between Twitter counts and citation counts of papers and (2) coverage of papers on Twitter.

(1) As the meta-analysis of Bornmann (2015) and further primary studies (e.g., de Winter, 2015; Jung, Lee, & Song, 2016) have shown, the correlation between tweets and citations is close to zero (see also Wouters et al., 2015). This result might mean that tweets are able to measure another dimension of research impact than citations. Two other possible interpretations are that (i) tweets are meaningless, since they correlate not even with citations. One reason for the possible missing meaning might be that tweets are very restricted in content (no more than 280 characters) (Haustein, Larivière, et al., 2014; Vainio & Holmberg, 2017). (ii) Tweets "influence science in indirect ways, for example by steering the popularity of research topics" (de Winter, 2015, p. 1776). The assumed meaninglessness of Twitter counts and the existence of possible indirect influencing ways have led to other analysis forms

---

[1] see https://help.altmetric.com/support/solutions/articles/6000060978-how-are-twitter-demographics

of Twitter data than counting tweets. In recent years, the following new approaches have been published going beyond simple tweets counting:

Costas, van Honk, Calero-Medina, and Zahedi (2017) produced thematic landscapes visualizing the topics on which people from a certain region tweet. The landscapes reveal, e.g., a strong health orientation in the thematic profile of African people. Robinson-Garcia, Arroyo-Machado, and Torres-Salinas (2019) also produced thematic landscapes: they used the overlay maps technique to identify topics of societal interest in microbiology. Hellsten and Leydesdorff (2018), Haunschild, Leydesdorff, and Bornmann (2019), and Haunschild, Leydesdorff, Bornmann, et al. (2019) produced Twitter networks mapping the co-occurrences of #hashtags, @usernames, and author keywords (of tweeted papers). These networks visualize public media-discussions on certain topics (reflecting the interacting connections between users and topics). Another Twitter network approach has been introduced by Robinson-Garcia, van Leeuwen, and Rafols (2016) that can be used for analyzing informal interactions between academics and their cities and revealing societal contributions of research. Sakaki, Okazaki, and Matsuo (2010) used Twitter data to produce a model that can "detect an earthquake with high probability (96% of earthquakes of Japan Meteorological Agency (JMA) seismic intensity scale 3 or more are detected)" (p. 851).

(2) As the literature overviews by Sugimoto et al. (2017) and Work, Haustein, Bowman, and Larivière (2015) show many studies have addressed the coverage of papers on Twitter (see also Hammarfelt, 2014; Thelwall & Kousha, 2015). These studies are important for using Twitter data in research evaluation, since this data source can only be reliably used if the coverage of papers is not too low. In other words, if 90% of the papers in a set did not receive any tweet, the data might be diagnostically less conclusive. Sugimoto et al. (2017) report that the coverage of papers on Twitter is around 10-20% – it depends on discipline, publication date, geographic regions, and other factors. Similar numbers can be found in Work et al. (2015). For example, the empirical study by Haustein, Peters, Sugimoto, Thelwall, and Larivière (2014) reports that "Twitter coverage at the discipline level is highest in Professional Fields, where 17.0% of PubMed documents were mentioned on Twitter at least once, followed by Psychology (14.9%) and Health (12.8%) … Twitter coverage is lowest for Physics papers covered by PubMed (1.8%)" (p. 662).

Some years ago, King et al. (2013) concluded (based on an empirical analysis of a dataset including tweets made about UK health reforms) that "to the best of our knowledge, there has been no analysis to date of how Twitter has been used to inform and debate a specific area of health policy" (p. 295). Our empirical results reported in section 4 and some studies from the overview in the following section show that the situation has changed.

**2.2    Spatial analysis of Twitter and other data on mental health**

In the previous section, we presented some studies using novel approaches based on network and mapping techniques to analyze Twitter data. These approaches are intended to reveal empirically public interactions with research. In this study, we would like to add another approach where Twitter data are spatially mapped and used as social-spatial sensors in the public health sector: we use location information in Twitter data to map the interest of Twitter users in publications on certain diseases. If people in regions which are concerned by high numbers of certain diseases tweet about these publications, it would mean that research reaches the people which should be reached (and not only other researchers who are interested in the results of colleagues). In other words, research would diffuse in practice and Twitter data used as social-spatial sensor would demonstrate that.

In this section, we review the relevant literature dealing with measuring online activities in the public health sector. Various researchers have already used Twitter data to "provide data about population-level health and behavior" (Sinnenberg et al., 2017).

According to Raghupathi and Raghupathi (2014), "big data in healthcare is overwhelming not only because of its volume but also because of the diversity of data types and the speed at which it must be managed". The data used in this research cover a broad spectrum from patient data in electronic patient records to data from the Web (e.g., Weblogs, Mei, Liu, Su, & Zhai, 2006, or Google) or social media (e.g., Twitter). Raghupathi and Raghupathi (2014) mention three areas where public health can profit from big data analyses: "1) analyzing disease patterns and tracking disease outbreaks and transmission to improve public health surveillance and speed response; 2) faster development of more accurately targeted vaccines, e.g., choosing the annual influenza strains; and, 3) turning large amounts of data into actionable information that can be used to identify needs, provide services, and predict and prevent crises, especially for the benefit of populations".

A very popular data source in recent years for analyzing diseases have been Google search queries. Either the studies analyzed Google web search logs directly or they used Google's web service Google Flu Trends (GFT) which aggregated Google search queries for making predictions about influenza activities (Nuti et al., 2014; see https://www.google.org/flutrends/about). For example, Ginsberg et al. (2009) analyzed individual searches of Google web search logs for tracking influenza-like illness in a population. The results show that "the relative frequency of certain queries is highly correlated with the percentage of physician visits in which a patient presents with influenza-like symptoms … This approach [tracking queries] may make it possible to use search queries to detect influenza epidemics in areas with a large population of web search users" (Ginsberg et al., 2009, p. 1012). In another study, Shaman, Karspeck, Yang, Tamerius, and Lipsitch (2013) used GFT data to develop a seasonal influenza prediction system. Various other studies – besides Ginsberg et al. (2009) and Shaman et al. (2013) – are reviewed in detail in the literature overview by Nuti et al. (2014). Their overview reveals that many studies have compared results based on GFT with external datasets to validate GFT results: "Over 90% of surveillance studies compared Google Trends with established data sets, which were often trusted sources of surveillance data. A large number of correlation studies had moderate to strong strengths of association, which demonstrates the potential of Google Trends data to be used for the surveillance of health-related phenomena" (Nuti et al., 2014).

Twitter data also have been used in health research (especially for public health predictions) "ranging from tracking infectious disease outbreaks, natural disasters, drug use, and more" (Pershad, Hangge, Albadawi, & Oklu, 2018). An overview of studies in this area can be found in Sinnenberg et al. (2017). Twitter data are especially interesting for health research, since the data contain meta-data on Twitter users (e.g., occupation or location) and tweets (e.g., timing or location). Most of the Twitter studies in health research "analyzed the content of tweets about a specific health topic to characterize public discourse on Twitter" (Sinnenberg et al., 2017) and compared the results with results based on external sources. According to Chunara, Andrews, and Brownstein (2012), Twitter data can be labeled as "informal" data since they are "unvetted by government or multilateral bodies such as the World Health Organization" (p. 39).

Signorini, Polgreen, and Segre (2010) analyzed public tweets that matched certain search terms (e.g., influenza and Tamiflu) and compared the results with influenza-like illness (ILI) values reported by the US Centers for Disease Control and Prevention (CDC). Their "regional model approximates the epidemic curve reported by ILI data with an average error of 0.37% (min=0.01%, max=1.25%) and a standard deviation of 0.26%" (Signorini et al., 2010). In a similar study, Chunara et al. (2012) assessed "correlation of volume of cholera-related HealthMap news media reports, Twitter postings, and government cholera cases reported in the first 100 days of the 2010 Haitian cholera outbreak. Trends in volume of informal sources significantly correlated in time with official case data and was available up to 2 weeks earlier" (p. 39). Lee, Agrawal, and Choudhary (2013) developed a real-time influenza and cancer surveillance system based on Twitter data (tweets which mention the words 'flu' or 'cancer') for tracking US influenza and cancer activities. The system might be very useful "not only for early prediction of seasonal disease outbreaks such as flu, but also for monitoring distribution of cancer patients with different cancer types and symptoms in each state and the popularity of treatments used" (Lee et al., 2013, p. 1474).

# 3 Methods

## 3.1 Dataset used

In this study, we use the diseases HIV, tuberculosis, and malaria as examples to demonstrate the use of Twitter data as social spatial sensors (see above). There are two open access websites that give worldwide data on all three diseases: World Bank Open Data (see https://data.worldbank.org) and WHO (see http://apps.who.int/gho/data). We used WHO data in this study because it is an internationally recognized organization. The WHO reports prevalence of HIV and incidence of tuberculosis and malaria. People who are infected with HIV can live fairly healthy lives with the disease but still infect others, and therefore prevalence (the proportion of cases in the population at a given time) is a better reporting measure than incidence (rate of occurrence of new cases). Incidence is a better reporting measure for diseases such as tuberculosis or malaria, because people either get treated and don't have the disease anymore, or they die.

We used three sources of data in this study: (1) incidence rates (or case numbers) for the three diseases (for countries or US states), (2) publications dealing with the diseases, and (3) tweets of these publications

### 3.1.1 Incidence rates (or case numbers)

For the HIV map, we used the data "Number of people (all ages) living with HIV; Estimates by country" (see http://apps.who.int/gho/data/view.main.22100?lang=en). The data are available for the years 2000, 2005, 2010, and 2018. Since Altmetric.com started to monitor Twitter in 2011, we used the mean of the national HIV cases from the years 2010 and 2018. In cases with no country data for 2010 or 2018, the existing value was used (either 2010 or 2018). The tuberculosis world map is based on WHO data retrieved from http://apps.who.int/gho/data/view.main.57040ALL?lang=en). The annual numbers of incident tuberculosis cases per country are available for the time period 2011-2017. Thus, we calculated the mean across the years for inclusion in the further statistical analysis. In cases, where the numbers are not available for all years, the mean was calculated based on the restricted set of years. For malaria, we applied a similar procedure as for tuberculosis: we used malaria incidences (per 1,000 population at risk) which are available for the years 2011-2017 and calculated the mean across the years. The malaria data are from http://apps.who.int/gho/data/node.main.MALARIAINCIDENCE?lang=en.

For HIV, we did not only produce a worldwide map, but also a state-specific US map. We would like to demonstrate that our mapping approach can be used very flexible (i.e., it can not only be used on the world level). The US data are from the CDC. We used the total number of cases from the table "Diagnoses of HIV infection, by area of residence" (Centers for Disease Control and Prevention, 2018, p. 114). Since the data are available for the years 2016 and 2017, we calculated the mean.

### 3.1.2 Publications

Publication sets regarding the diseases were downloaded from PubMed. The following search queries were used (see Baumann, 2016):

1) HIV-related papers: ("hiv"[MeSH Major Topic]) AND ("2011/01/01"[Date - Publication] : "2017/12/31"[Date - Publication])

2) Tuberculosis-related papers: ("tuberculosis"[MeSH Major Topic]) AND ("2011/01/01"[Date - Publication] : "2017/12/31"[Date - Publication])

3) Malaria-related papers: ("malaria"[MeSH Major Topic]) AND ("2011/01/01"[Date - Publication] : "2017/12/31"[Date - Publication])

We downloaded the PubMedIDs for the papers found using the aforementioned search queries on 27 January 2020. In total, we downloaded 17,295 PubMedIDs from papers regarding HIV, 26,595 PubMedIDs from papers regarding tuberculosis, and 13,974 PubMedIDs from papers regarding malaria.

### 3.1.3 Tweets

The PubMedIDs were imported into our in-house PostgreSQL database and matched with the Twitter data from Altmetric.com via the PubMedID. 8,442 of the HIV-related papers (48.8%) were tweeted by in total 55,506 tweets. 11,139 of the tuberculosis-related papers (41.9%) were tweeted by in total 85,737 tweets. 8,403 of the malaria-related papers (60.1%) were tweeted by in total 73,111 tweets. Overall, the papers regarding these three diseases have been mentioned in tweets more often than medical papers in general (see Haustein, Peters, et al., 2014).

The Tweet ID was exported from the Altmetric database. The three sets of tweets were downloaded via the Twitter API using R (R Core Team, 2019) and stored in local SQLite database files using the R package RSQLite (Müller, Wickham, James, & Falcon, 2017). Functions from the R package DBI (R Special Interest Group on Databases (R-SIG-DB), Wickham, & Müller, 2018) were used for sending database queries. Not all Twitter users provide information about their geographical location (Sakaki et al., 2010; Wouters, Zahedi, & Costas, 2019). From the tweets mentioning HIV-related papers, only 342 tweets contained precise geo-coordinates, but 39,967 of those tweets contained some free-text user location information. In the case of the tweets mentioning tuberculosis-related papers, only 83 tweets contained precise geo-coordinates, but 62,730 of those tweets contained some free-text user location information. In the case of tweets mentioning malaria-related papers, only 85 tweets contained precise geo-coordinates, but 69,420 of those tweets contained some free-text user location information. We discarded the precise geo-coordinates and used only the user location information.

One problem with the free-text user location information is that some users seem to become very imaginative. In order to reduce wrong location information, we needed to filter the location information for meaningful entries. We imported the city and country names from the Global Research Identifier Database (GRID, https://grid.ac/) for obtaining a whitelist of existing cities and countries. Only that location information was kept which contained a city name and a country name from the GRID database. Usage of the GRID database introduces a potential bias towards city names with research institutes. For the countries, we added "USA" and "UK" (requiring a comma or whitespace before "UK"). In both data sets, location information and GRID data, we removed non-standard characters (e.g., ö, ä, ß, ê) by only keeping the standard alphabet letters, whitespaces, and some punctuation characters (i.e., "," and ";"). In addition, both data sets were converted into lower case characters for matching.

We found spurious location strings by manual inspection. Location strings which contained one of the following strings were excluded: "www", "http", "not from", "worldwide", "everywhere", "mostly nucleus", "bcnvcia", "&", " and ", " und ", and " y ". The latter four exclusion strings are expected to remove multiple locations in a single location string (e.g., "Washington DC & New Delhi"). The remaining location strings were passed to the Google API via the R package ggmap (Kahle & Wickham, 2013) if the location strings contained more than three characters. The Google API returned among others precise geo-coordinates, country, and state names (if available) which were stored in a CSV file for plotting and statistical analysis. For the tweets mentioning HIV-related papers, we obtained by this procedure 10,018 geo-coordinates (18.1% of all tweets), for the tweets mentioning tuberculosis-related papers 16,966 geo-coordinates (19.8% of all tweets), and for the tweets mentioning malaria-related papers 13,780 geo-coordinates (18.9% of all tweets). An example R script is available at http://ivs.fkf.mpg.de/twitter_maps/get_location_information_from_tweet_ids.R. We also used the GRID database with Dimensions' publication data (Herzog, Hook, & Konkiel, 2020) shared by Digital Science with us to count the papers of author country and state (in the case of USA for the HIV-related papers) codes.

It should be considered in the interpretation of the Twitter data that censorship of Twitter in certain countries exists which "refers to Internet censorship by governments that block access to Twitter, or censorship by Twitter itself" (https://en.wikipedia.org/wiki/Censorship_of_Twitter).

### 3.2  Statistics applied

We used several Stata commands to produce the social-spatial Twitter maps (Crow & Gould, 2013; Huebler, 2012; StataCorp., 2017). The most important Stata commands were `shp2dta` (Crow, 2006) and `spmap` (Pisati, 2007). We additionally calculated Poisson regression models with number of tweets as dependent variable and number of disease cases (e.g., HIV cases) and number of papers as independent variables. Poisson regression models are indicated with count variables as dependent variables (Deschacht & Engels, 2014; Hilbe, 2014). In the interpretation of the models, we focus on percentage changes in expected counts (Long & Freese, 2014). These percentages show for a standard deviation increase in the number of disease cases in a country (or state), the increases in the expected tweet number in that country (or state), holding the country's (or state's) number of papers constant.

## 4    Results

In section 2, we reviewed the literature using internet data (Twitter data) in health research. One important outcome of these and similar studies was "the surveillance of influenza outbreaks with comparable accuracy to traditional methodologies" (Nuti et al., 2014). The results might indicate that Twitter activity reflects the interest of the general public in research findings. The results confirm the statement by Robinson-Garcia et al. (2019) that by tracking alternative channels "it is possible to identify and access literature which might not only be relevant to scientists, but also to lay people".

On a related note, Sakaki et al. (2010) coined the term "social sensor" which means that tweets are regarded as sensory information and Twitter users as sensors. The use and interpretation of Twitter activity as social sensors has not been done in altmetrics research hitherto. The activity of Twitter sensors – which can be in the status "active" (i.e., tweeting) or not – on certain triggers (e.g., earth quakes or indications of influenza) can be measured. In this study, Twitter users function as social-spatial sensors by being aware of papers dealing with a certain disease. Since one can expect that the interest in papers on certain diseases increases, when the user is located in regions with many cases of illness, Twitter rates and disease rates might correlate.

Using Twitter data as sensors, we investigate in this study whether research on a certain disease (tuberculosis, malaria, and HIV) reaches the people that are especially concerned by the disease (regions with many people having the disease). Since Twitter data contain location information, which can be converted into geo-coordinates (see section 3), it is possible to map Twitter activity on certain papers (e.g., dealing with tuberculosis, malaria, or HIV). Each tweet is represented by a single dot on the map.

## 4.1  Mapping tuberculosis related data

Figure 1 shows worldwide Twitter activity referring to papers dealing with tuberculosis. The underlying blue-colored scheme visualizes the number of incident tuberculosis cases per country. The map is intended to show whether tuberculosis research reaches regions with many tuberculosis cases: does the number of tuberculosis cases correlate with the number of tweets on tuberculosis papers?

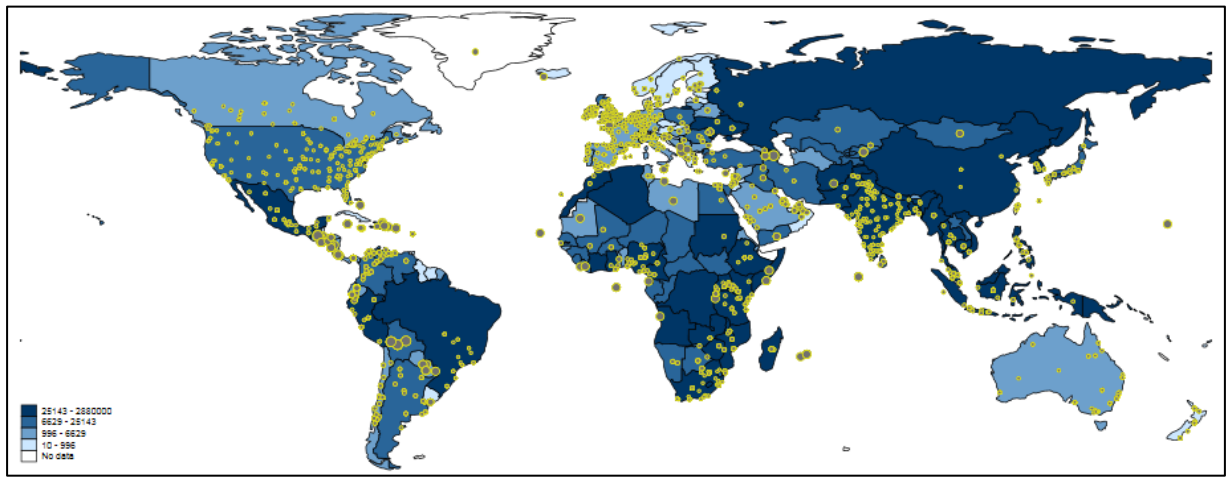

Figure 1. Tweeting on papers dealing with tuberculosis worldwide. Each tweet is inversely weighted with the number of papers published by authors in the corresponding country: the larger the dots, the smaller the research activity. The countries are colored according to the total number of incident tuberculosis cases. For some countries, e.g. Greenland, no data are available. Some countries such as China or Iran block internet access to Twitter or its content (see section 3.1.3).

One of the problems with Twitter data in the context of this study is that Twitter activity is generally high where much research is done (see, e.g., Western Europe or the Boston area in Figure 1). Since this is not the activity which we intended to measure, we inversely weighted the size of each tweet on the map by the number of papers in that country [i.e., 1/log(number of papers)]. Thus, the more papers' authors are located in a country, the smaller the size of the tweet dot is (see here Ginsberg et al., 2009; Sinnenberg et al., 2017). We assume that large dots reflect tweets of people not doing research or not being a publisher/ publishing organization (but might be personally confronted with tuberculosis).

The map in Figure 1 might show the expected result that high Twitter activity is related to high numbers of incident tuberculosis cases. However, it is not completely clear whether this conclusion can be drawn, since there are several countries with high Twitter activity and high paper output (e.g., Western Europe and the Boston region). For some regions on the map, the extent of Twitter activity is difficult to interpret since tweet dots might overlap (especially those with larger sizes). To have a conclusive answer on the relation between Twitter activity and paper output, we additionally calculated Poisson regression models with number of tweets as dependent variable and number of incident tuberculosis cases and number of papers as independent variables.

The results are shown in Table 1. The coefficients of both independent variables are statistically significant. The percentage changes in expected counts reveal that incident tuberculosis cases and Twitter activities are related in fact: for a standard deviation increase in the number of incident tuberculosis cases in a country, the expected number of tweets in that state increases by 7.9%, holding the country's number of papers constant. The results in Table 1 further show that the influence of the number of incident tuberculosis cases is significantly smaller than that of the number of papers. This might reveal the stronger dependency of Twitter data from the science sector than the general public (people concerned by the disease).

Table 1. Coefficients of a Poisson regression model with number of tweets as dependent variable (n=126 countries)

| Independent variable | Coefficient | Standard error | Percentage change in expected count |
|---|---|---|---|
| Number of incident tuberculosis cases | $2.63 \times 10^{-7}$*** | $1.08 \times 10^{-8}$ | 7.9 |
| Number of papers | $7.69 \times 10^{-4}$*** | $3.72 \times 10^{-6}$ | 61.9 |
| Constant | 4.31*** | .01 | |

Notes. *** $p<.001$

## 4.2  Mapping malaria related data

The map visualizing Twitter activity as social-spatial sensor of the public use of malaria-related literature is shown in Figure 2. The blue coloring of the countries reflects malaria incidences (per 1,000 population at risk) from the WHO (see above). Since the WHO malaria data does not include all countries worldwide, many countries are white-colored. The countries with available data are concentrated on South America, Africa, and Asia. Whereas some countries in Africa are characterized by high Twitter activity and high incidence rates (e.g., Ghana), other countries (e.g., Chad) have high incidence rates but not any Twitter activity.

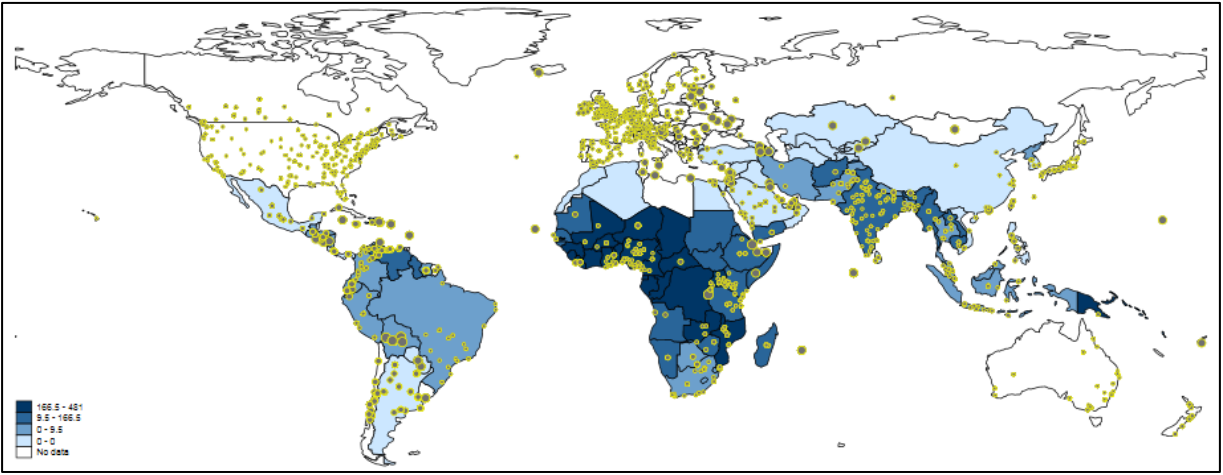

Figure 2. Tweeting on papers dealing with malaria worldwide. Each tweet is inversely weighted with the number of papers published by authors in the corresponding country: the larger the dots, the smaller the research activity. The countries are colored according to malaria incidences (per 1,000 population at risk). For various countries, e.g. Russia and Australia, no data are available. Some countries such as China or Iran block internet access to Twitter or its content (see section 3.1.3).

Following the statistical analysis in section 4.1, we additionally calculated a Poisson regression model to investigate the relationship of incidence rates and Twitter activity in more detail. The data from 79 countries could be considered in the regression model (these are the countries with available data for the three variables). The results are reported in Table 2. They point out that both independent variables – number of papers and malaria incidences – are statistically significant.

Table 2. Coefficients of a Poisson regression model with number of tweets as dependent variable (n=79 countries)

| Independent variable | Coefficient | Standard error | Percentage change in expected count |
| --- | --- | --- | --- |
| Malaria incidences (per 1,000 population at risk | $1.24 \times 10^{-3}$*** | $1.16 \times 10^{-4}$ | 17.1 |
| Number of papers | $3.54 \times 10^{-3}$*** | $5.02 \times 10^{-5}$ | 86.3 |
| Constant | 3.16*** | .03 | |

Note. *** p<.001

The relevant information in Table 2 for interpreting the results of the regression analysis are the percentage changes in expected counts. These results reveal that malaria incidences and Twitter activities are related in fact: for a standard deviation increase in malaria incidences in a country, the expected number of tweets in that state increases by 17.1%, holding the country's number of papers constant. Although the publication numbers also have a substantial influence on the Twitter activity (the percentage change in expected counts is 86.3%), Twitter activity seems to reflect the use of papers on malaria in affected regions.

**4.3    Mapping HIV related data**

Figure 3 shows the HIV world map. The countries are colored using the total number of HIV cases (see above). For several countries, no data are available, e.g., Switzerland and Canada. The number of tweets seems to be related to the national number of HIV cases. There are, however, many countries with relatively high numbers of HIV cases, but without any Twitter activity. Low-income countries share the highest burden of HIV cases but countries such as Niger, Chad, Sudan, and Central African Republic don't have Twitter activity. This may reflect inadequate access to the Twitter platform because of a lack of computers, mobile devices, and internet. In addition, health research in low and middle-income countries is insufficient and fragmented, although it is critical for overcoming global health challenges.

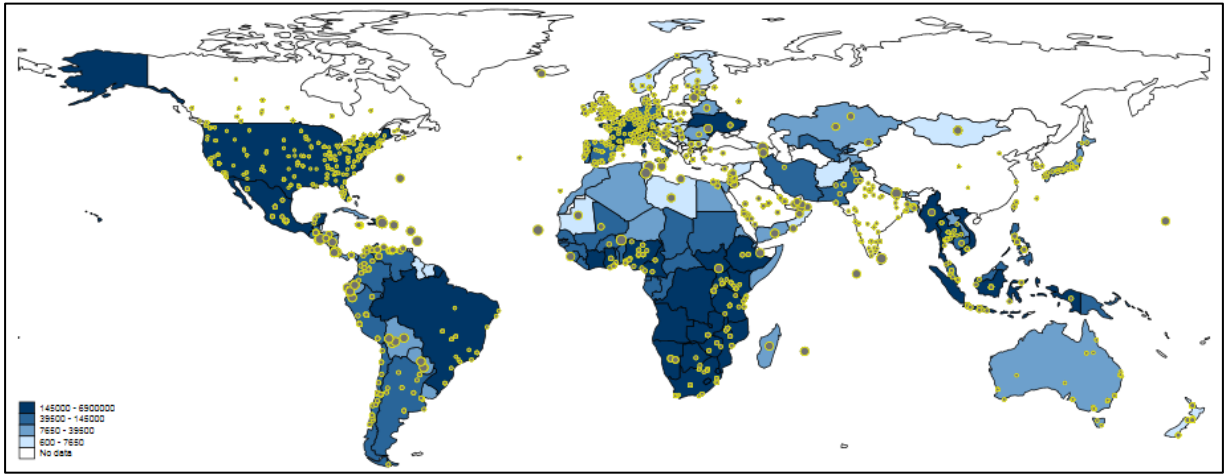

Figure 3. Tweeting on papers dealing with HIV worldwide. Each tweet is inversely weighted with the number of papers published by authors in the corresponding country: the larger the dots, the smaller the research activity. The countries are colored according to the total number of HIV cases. For some countries, e.g., Switzerland, no data are available. Some countries such as China or Iran block internet access to Twitter or its content (see section 3.1.3).

The results of the Poisson regression analysis are shown in Table 3. Only 89 countries could be considered in the analysis, since only countries have been included with available data for all included variables. The percentage changes in expected counts reveal that there is a relationship between HIV cases and Twitter activities: for a standard deviation increase in the number of HIV cases in a country, the expected number of tweets in that country increases by 22.3%, holding the country's number of papers constant.

Table 3. Coefficients of a Poisson regression model with number of tweets as dependent variable (n=89 countries)

| Independent variable | Coefficient | Standard error | Percentage change in expected count |
|---|---|---|---|
| Number of HIV cases | $2.45 \times 10^{-7}$*** | $8.44 \times 10^{-9}$ | 22.3 |
| Number of papers | $4.76 \times 10^{-4}$*** | $3.34 \times 10^{-6}$ | 50.5 |
| Constant | 3.73*** | .02 | |

Note. *** $p<.001$

Twitter data cannot only be used as social sensors on the country level, but can be restricted to a single country. We would like to demonstrate the approach based on US HIV-related data. Figure 4 shows paper-based Twitter activity dealing with HIV in the USA. The blue-colored scheme presents the number of HIV cases per US state. The map might show that the numbers of HIV cases in the US states are in fact related to the number of tweets on HIV papers. However, there are several US states with high Twitter activity and high paper output (e.g., the Boston region).

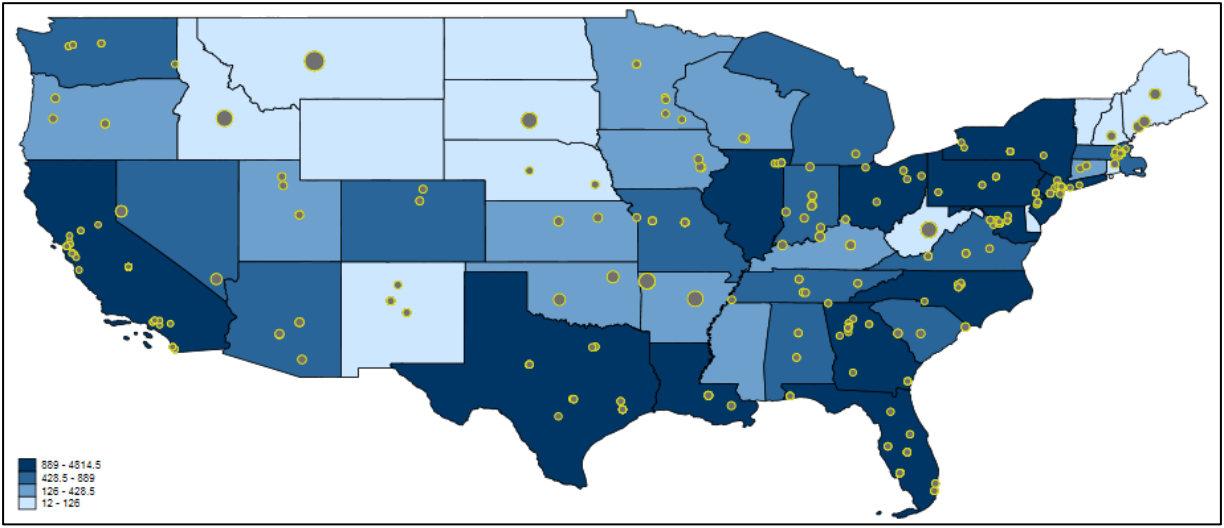

Figure 4. Tweeting on papers dealing with HIV in the USA. Each tweet is inversely weighted with the number of papers published by authors in the corresponding US state: the larger the dots, the smaller the research activity. The US states are colored according to the total number of HIV cases in 2016/2017 (Centers for Disease Control and Prevention, 2018).

We calculated Poisson regression models with number of HIV cases and number of papers as independent variables and number of tweets as dependent variable. Table 4 reports the results. The results are based on a reduced number of US states (44 instead of 51) since only US states are considered with at least one tweet. The percentage changes in expected counts in Table 4 point out that HIV cases and Twitter activities seem to be correlated: for a standard deviation increase in the number of HIV cases in a US state, the expected number of tweets in that state increases by 69.6%, holding the US state's number of papers constant. The results in Table 4 further show that the influence of the number of HIV cases is greater than that of the number of papers. This is different from the worldwide results (see above) where the number of papers' influence exceeds the influence of Twitter activity. In the US states, there is a stronger dependency of Twitter data from the general public (people concerned) than from the science sector.

Table 4. Coefficients of Poisson regression model with number of tweets as dependent variable (n=44 US states)

| Independent variable | Coefficient | Standard error | Percentage change in expected count |
|---|---|---|---|
| Number of HIV cases | $4.40 \times 10^{-4}$*** | $1.52 \times 10^{-5}$ | 69.6 |
| Number of papers | $6.391 \times 10^{-4}$*** | $4.08 \times 10^{-5}$ | 33.4 |
| Constant | 2.70*** | .04 | |

Note. *** $p<.001$

# 5 Discussion

The use of altmetrics and especially Twitter data in research evaluation (impact measurements) has been assessed very critically in the past years (Tahamtan & Bornmann, 2020; Zahedi, Costas, & Wouters, 2014). The empirical results by Robinson-Garcia, Ramos-Vielba, Costas, D'Este, and Rafols (2017) demonstrate "an absence of relation between altmetric coverage of researchers and the number of types of non-academic partners with whom they interact". According to Haustein et al. (2016), "systematic evidence as to whether tweets are valid markers of actual societal and/or scientific impact is lacking" (p. 233). Thelwall and Kousha (2015) reviewed the literature on various Web indicators and altmetrics and concluded: "only Google Patents citations and clinical guideline citations clearly reflect wider societal impact and no social media metrics do" (p. 615). Most of the empirical studies in the past focused on using Twitter counts in research evaluation.

In this study, we propose to abstain from using Twitter data as simple counts, but to use additional meta-data, which are accessible for single tweets and Twitter users. For Pershad et al. (2018), "through its ability to connect millions of people with public tweets,

Twitter has the potential to revolutionize public health efforts, including disseminating health updates, sharing information about diseases, or coordinating relief efforts". The effects of these activities could be measured by hashtag or user networks or spatial maps based on Twitter data. The use of Twitter data is especially helpful in countries or regions "where conventional data collection may be challenging and resource intensive" (Nuti et al., 2014). In this paper, we propose to use Twitter as a supplement to Google web search logs which have been already used as data source for a broad-reaching influenza monitoring system: "whereas traditional systems require 1–2 weeks to gather and process surveillance data, our estimates are current each day" (Ginsberg et al., 2009, p. 1014). Google web searches have been seen as an attractive data source for empirical research in the past, since people search for diseases, symptoms, and medical treatments (Lee et al., 2013). Fung et al. (2017) analyzed the online discussion of major diseases (including tuberculosis, malaria, and HIV) using hashtags. They, however, did not study the relation to scientific papers regarding these diseases.

Science mapping has become an important method in research evaluation (Robinson-Garcia et al., 2019). The literature reviewed in section 2.2 demonstrated that Twitter data are well-suited for science mapping activities; according to Kuehn (2015), they are especially interesting for the health care area: they have "the potential to provide early warnings about chronic disease, emergencies, adverse drug reactions, or even safety problems like prescription drug misuse" (p. 2010). In this study, we propose to use Twitter data as social-spatial sensors. We are interested in the question whether research papers on certain diseases (tuberculosis, malaria, and HIV) are perceived by people in regions (worldwide) which are especially concerned by the diseases. We used two methods for answering this question: (1) we visualized meta-data of tweets that include links to disease-related research papers in combination with spatial maps reflecting incidence rates or number of disease cases. It can be assessed by visual inspection then whether Twitter activity is related to incidence rates or number of cases. (2) We used regression models to analyze the relationship between Twitter activity and incidence rates or number of cases. In these models, the number of papers has been controlled to consider that Twitter activity depends on research activity.

The results of the social-spatial Twitter maps and regression models reveal that the combination of both methods is useful to answer our research question. We received an impression of how research papers on tuberculosis, malaria, and HIV have been perceived by people in regions which are especially concerned by the diseases. The maps give a spatial overview and a first impression; the regression models quantify the attention research papers have received regionally. For example, the comparison of the regression model results for tuberculosis, malaria, and HIV reveal that research papers might be of specific public interest with respect to HIV. The percentage in expected counts is higher than those for malaria and tuberculosis. The HIV Twitter analysis that focused on US states shows that the relationship between public attention and number of HIV cases is higher than the relationship between public attention and research activity. The results might suggest how important social media platforms are in diffusing research into areas where diseases are more prevalent but research outputs are low.

Our study might demonstrate an interesting approach for using Twitter data for research evaluation purposes. The data should be used, however, with care. A first important point concerns the location information from Twitter. Wouters et al. (2019) identified two challenges in using such data: "1) the lack of disclosure of geographic information for all social media users (e. g., not all users in Mendeley, Facebook or Twitter disclose their geo-location), and 2) the variable granularity of available geographic information (e. g., not all users disclose their full geographical information; some provide only country-level information, while others also disclose region or location)" (p. 701). A second point refers to the construct validity of the Twitter data as used in this study (see here also Bornmann, Haunschild, & Adams, 2019). Construct validity is "the degree to which a test measure (e.g. a reference count or an Altmetric score) measures what it claims or purports to be measuring (e.g. quality or social engagement)?" (Rowlands, 2018). We do not know whether tweets really reflect the usefulness of research papers (on certain diseases). We assume that based on the relationship between Twitter activity and incidence rates or case numbers of certain diseases. However, there might be other factors relevant for explaining this relationship (e.g., restricted access to Twitter or censorship by Twitter).

A third point refers to the content of tweets: can we assume that tweets deal with the "right" papers which are helpful with respect to certain diseases? Other papers might be more helpful. Furthermore, we cannot assume in every case that "true stories" on papers are distributed on Twitter. The results by Vosoughi, Roy, and Aral (2018) show that false news stories are frequently distributed on Twitter and that "falsehood diffused significantly farther, faster, deeper, and more broadly than the truth in all categories of information". Pershad et al. (2018) point to the result of a study that found that about 20% of tweets about healthcare contained inaccurate information. Robinson-Garcia, Costas, Isett, Melkers, and Hicks (2017) list some proposals that can improve the data quality of tweets (e.g., by removing the data from certain accounts which have been identified as problematic hitherto). Future studies might show whether the consideration of these proposals leads to other (better) results than those presented in this paper.

We recommend that our proposal of using Twitter data should be tested by other research groups (active in scientometrics). According to Thelwall (2017), "indicators must be evaluated before they can be used with any confidence. Evaluations can assess the type of impact represented by the indicator and the strength of the evidence that it provides". Beyond testing our approach, future studies could investigate whether papers synthesizing research (or papers close to clinical practice) are more popular on Twitter than papers reporting results from basic research (see here Andersen & Haustein, 2015). Future studies also could have a more focused view on certain drugs, therapies or prophylaxis by investigating their reflections in Twitter activity. Another idea is to focus on papers publishing research funded by a special organization. Two organizations could be compared: which organization is better able to reach target groups than the other?

In this study, we focus on the health care sector to demonstrate our proposal of using Twitter data as social-spatial sensors. Our proposal, however, is not only restricted to this sector. Sakaki et al. (2010) list some possible other sectors in which our proposal might be able to be applied (e.g., natural events such as climate change consequences).

# Acknowledgements

The twitter data are retrieved from our locally maintained database at the Max Planck Institute for Solid State Research (MPI-FKF, Stuttgart) and derived from data shared with us by the company Altmetric.com on October 30, 2019. Tweets with their location information were retrieved from the Twitter API. The authors thank Rodrigo Costas (CWTS) and Stacy Konkiel (Altmetric.com) for helpful discussions regarding the analysis of location information of Twitter users.

# References

Andersen, J. P., & Haustein, S. (2015). Influence of study type on Twitter activity for medical research papers. In A. A. Salah, Y. Tonta, A. A. A. Salah, C. Sugimoto & U. Al (Eds.), *The 15th Conference of the International Society for Scientometrics and Informetrics* (pp. 26-36). Istanbul, Turkey: ISSI, Boaziçi University Printhouse.

Baumann, N. (2016). How to use the medical subject headings (MeSH). *International Journal of Clinical Practice, 70*(2), 171-174. doi: 10.1111/ijcp.12767.

Bik, H. M., & Goldstein, M. C. (2013). An introduction to social media for scientists. *PLoS Biol, 11*(4), e1001535. doi: 10.1371/journal.pbio.1001535.

Blümel, C., Gauch, S., & Beng, F. (2017). Altmetrics and its intellectual predecessors: Patterns of argumentation and conceptual development. In P. Larédo (Ed.), *Proceedings of the Science, Technology, & Innovation Indicators Conference "Open indicators: innovation, participation and actor-based STI indicators*. Paris, France.

Bornmann, L. (2015). Alternative metrics in scientometrics: A meta-analysis of research into three altmetrics. *Scientometrics, 103*(3), 1123-1144.

Bornmann, L. (2016). Scientific revolution in scientometrics: The broadening of impact from citation to societal. In C. R. Sugimoto (Ed.), *Theories of informetrics and scholarly communication* (pp. 347-359). Berlin, Germany: De Gruyter.

Bornmann, L., & Haunschild, R. (2017). Does evaluative scientometrics lose its main focus on scientific quality by the new orientation towards societal impact? *Scientometrics, 110*(2), 937-943. doi: 10.1007/s11192-016-2200-2.

Bornmann, L., Haunschild, R., & Adams, J. (2019). Do altmetrics assess societal impact in a comparable way to case studies? An empirical test of the convergent validity of altmetrics based on data from the UK research excellence framework (REF). *Journal of Informetrics, 13*(1), 325-340. doi: 10.1016/j.joi.2019.01.008.

Centers for Disease Control and Prevention. (2018). HIV Surveillance Report, 2017 (vol. 29). Retrieved October 30, 2019, from http://www.cdc.gov/hiv/library/reports/hiv-surveillance.html

Chunara, R., Andrews, J. R., & Brownstein, J. S. (2012). Social and news media enable estimation of epidemiological patterns early in the 2010 Haitian cholera outbreak. *American Journal of Tropical Medicine and Hygiene, 86*(1), 39-45. doi: 10.4269/ajtmh.2012.11-0597.

Colledge, L. (2014). *Snowball metrics recipe book*. Amsterdam, the Netherlands: Snowball Metrics program partners.

Costas, R., van Honk, J., Calero-Medina, C., & Zahedi, Z. (2017). Exploring the descriptive power of altmetrics: Case study of Africa, USA and EU28 countries (2012-2014). In P. Larédo (Ed.), *Proceedings of the Science, Technology, & Innovation Indicators Conference "Open indicators: innovation, participation and actor-based STI indicators*. Paris, France.

Crow, K. (2006). SHP2DTA: Stata module to converts shape boundary files to Stata datasets, Statistical Software Components S456718, Boston College Department of Economics, revised 17 Jul 2015.

Crow, K., & Gould, W. (2013). Working with spmap and maps. Retrieved February 10, 2020, from https://www.stata.com/support/faqs/graphics/spmap-and-maps/

de Winter, J. C. F. (2015). The relationship between tweets, citations, and article views for PLOS ONE articles. *Scientometrics, 102*(2), 1773-1779. doi: 10.1007/s11192-014-1445-x.

Deschacht, N., & Engels, T. E. (2014). Limited dependent variable models and probabilistic prediction in informetrics. In Y. Ding, R. Rousseau & D. Wolfram (Eds.), *Measuring scholarly impact* (pp. 193-214): Springer International Publishing.

Erdt, M., Nagarajan, A., Sin, S.-C. J., & Theng, Y.-L. (2016). Altmetrics: An analysis of the state-of-the-art in measuring research impact on social media. *Scientometrics, 109*, 1117–1166. doi: 10.1007/s11192-016-2077-0.

Fung, I. C. H., Jackson, A. M., Ahweyevu, J. O., Grizzle, J. H., Yin, J. J., Tse, Z. T. H., . . . Fu, K. W. (2017). #Globalhealth Twitter Conversations on #Malaria, #HIV, #TB, #NCDS, and #NTDS: a Cross-Sectional Analysis. *Annals of Global Health, 83*(3-4), 682-690. doi: 10.1016/j.aogh.2017.09.006.

Ginsberg, J., Mohebbi, M. H., Patel, R. S., Brammer, L., Smolinski, M. S., & Brilliant, L. (2009). Detecting influenza epidemics using search engine query data. *Nature, 457*(7232), 1012-U1014. doi: 10.1038/nature07634.

González-Valiente, C. L., Pacheco-Mendoza, J., & Arencibia-Jorge, R. (2016). A review of altmetrics as an emerging discipline for research evaluation. *Learned Publishing, 29*(4), 229-238. doi: 10.1002/leap.1043.

Hammarfelt, B. (2014). Using altmetrics for assessing research impact in the humanities. *Scientometrics*, 1-12. doi: 10.1007/s11192-014-1261-3.

Haunschild, R., Leydesdorff, L., & Bornmann, L. (2019). *Library and Information Science papers as Topics on Twitter: A network approach to measuring public attention*. Paper presented at the ISSI 2019 – 17th International Conference of the International Society for Scientometrics and Informetrics, Rome, Italy.

Haunschild, R., Leydesdorff, L., Bornmann, L., Hellsten, I., & Marx, W. (2019). Does the public discuss other topics on climate change than researchers? A comparison of networks based on author keywords and hashtags. *Journal of Informetrics, 13*(2), 695-707.

Haustein, S. (2016). Grand challenges in altmetrics: heterogeneity, data quality and dependencies. *Scientometrics, 108*(1), 413-423. doi: 10.1007/s11192-016-1910-9.

Haustein, S. (2019). Scholarly Twitter Metrics. In W. Glänzel, H. F. Moed, U. Schmoch & M. Thelwall (Eds.), *Springer Handbook of Science and Technology Indicators* (pp. 729-760). Cham, Switzerland: Springer International Publishing.

Haustein, S., Bowman, T. D., Holmberg, K., Tsou, A., Sugimoto, C. R., & Larivière, V. (2016). Tweets as impact indicators: Examining the implications of automated bot accounts on Twitter. *Journal of the Association for Information Science and Technology, 67*(1), 232-238.

Haustein, S., Larivière, V., Thelwall, M., Amyot, D., & Peters, I. (2014). Tweets vs. Mendeley readers: How do these two social media metrics differ? *it – Information Technology, 56*(5), 207-215.

Haustein, S., Peters, I., Sugimoto, C. R., Thelwall, M., & Larivière, V. (2014). Tweeting biomedicine: An analysis of tweets and citations in the biomedical literature. *Journal of the Association for Information Science and Technology, 65*(4), 656-669. doi: 10.1002/asi.23101.

Hellsten, I., & Leydesdorff, L. (2018). Automated analysis of topic-actor networks on Twitter: New approach to the analysis of socio-semantic networks. *Journal of the Association for Information Science and Technology, 71*(1), 3-15.

Herzog, C., Hook, D., & Konkiel, S. (2020). Dimensions: Bringing down barriers between scientometricians and data. *Quantitative Science Studies, 1*(1), 387-395. doi: 10.1162/qss_a_00020.

Hilbe, J. M. (2014). *Modelling count data*. New York, NY, USA: Cambridge University Press.

Huebler, F. (2012). Guide to creating maps with Stata. Retrieved February 10, 2020, from https://huebler.blogspot.com/2012/08/stata-maps.html

Jung, H., Lee, K., & Song, M. (2016). Examining characteristics of traditional and Twitter citation. *Frontiers in Research Metrics and Analytics, 1*(6). doi: 10.3389/frma.2016.00006.

Kahle, D., & Wickham, H. (2013). ggmap: Spatial Visualization with ggplot2. *The R Journal, 5*(1), 144-161.

King, D., Ramirez-Cano, D., Greaves, F., Vlaev, I., Beales, S., & Darzi, A. (2013). Twitter and the health reforms in the English National Health Service. *Health Policy, 110*(2-3), 291-297. doi: 10.1016/j.healthpol.2013.02.005.

Konkiel, S., Madjarevic, N., & Rees, A. (2016). Altmetrics for Librarians: 100+ tips, tricks, and examples, from http://dx.doi.org/10.6084/m9.figshare.3749838

Kuehn, B. M. (2015). Twitter streams fuel big data approaches to health forecasting. *Journal of the American Medical Association, 314*(19), 2010-2012. doi: 10.1001/jama.2015.12836.

Lee, K., Agrawal, A., & Choudhary, A. (2013). *Real-time disease surveillance using twitter data: Demonstration on flu and cancer.* Paper presented at the Proceedings of the 19th ACM SIGKDD international conference on knowledge discovery and data mining.

Long, J. S., & Freese, J. (2014). *Regression models for categorical dependent variables using Stata* (3. ed.). College Station, TX, USA: Stata Press, Stata Corporation.

Mas-Bleda, A., & Thelwall, M. (2016). Can alternative indicators overcome language biases in citation counts? A comparison of Spanish and UK research. *Scientometrics, 109*(3), 2007-2030. doi: 10.1007/s11192-016-2118-8.

Mei, Q., Liu, C., Su, H., & Zhai, C. (2006). A probabilistic approach to spatiotemporal theme pattern mining on weblogs. Retrieved February 13, 2010, from http://www-personal.umich.edu/~qmei/pub/www06-blog.pdf

Moed, H. F. (2017). *Applied Evaluative Informetrics*. Heidelberg, Germany: Springer.

Müller, K., Wickham, H., James, D. A., & Falcon, S. (2017). RSQLite: 'SQLite' Interface for R. R package version 2.0, from https://CRAN.R-project.org/package=RSQLite

Nuti, S. V., Wayda, B., Ranasinghe, I., Wang, S., Dreyer, R. P., Chen, S. I., & Murugiah, K. (2014). The use of Google Trends in health care research: A systematic review. *PLOS ONE, 9*(10), e109583. doi: 10.1371/journal.pone.0109583.

Pershad, Y., Hangge, P. T., Albadawi, H., & Oklu, R. (2018). Social medicine: Twitter in healthcare. *Journal of Clinical Medicine, 7*(6), 121. doi: 10.3390/jcm7060121.

Pisati, M. (2007). *SPMAP: Stata module to visualize spatial data, Statistical Software Components S456812, Boston College Department of Economics, revised 18 Jan 2018*.

Priem, J., & Costello, K. L. (2010). How and why scholars cite on Twitter. *Proceedings of the American Society for Information Science and Technology, 47*(1), 1-4. doi: 10.1002/meet.14504701201.

R Core Team. (2019). R: A Language and Environment for Statistical Computing (Version 3.6.0). Vienna, Austria: R Foundation for Statistical Computing. Retrieved from https://www.r-project.org/

R Special Interest Group on Databases (R-SIG-DB), Wickham, H., & Müller, K. (2018). DBI: R Database Interface.

Raghupathi, W., & Raghupathi, V. (2014). Big data analytics in healthcare: promise and potential. *Health Information Science and Systems, 2*(1), 3. doi: 10.1186/2047-2501-2-3.

Robinson-Garcia, N., Arroyo-Machado, W., & Torres-Salinas, D. (2019). Mapping social media attention in Microbiology: identifying main topics and actors. *FEMS Microbiology Letters, 366*(7). doi: 10.1093/femsle/fnz075.

Robinson-Garcia, N., Costas, R., Isett, K., Melkers, J., & Hicks, D. (2017). The unbearable emptiness of tweeting—About journal articles. *PLOS ONE, 12*(8), e0183551. doi: 10.1371/journal.pone.0183551.

Robinson-Garcia, N., Ramos-Vielba, I., Costas, R., D'Este, P., & Rafols, I. (2017). Do altmetric indicators capture societal engagement? A comparison between survey and social media data. In P. Larédo (Ed.), *Proceedings of the Science, Technology, & Innovation Indicators Conference "Open indicators: innovation, participation and actor-based STI indicators*. Paris, France.

Robinson-Garcia, N., van Leeuwen, T. N., & Rafols, I. (2016). SSH & the city. A network approach for tracing the societal contribution of the social sciences and humanities for local development. In I. Ràfols, J. Molas-Gallart, E. Castro-Martínez & R. Woolley (Eds.), *Proceedings of the 21 ST International Conference on Science and Technology Indicator*. València, Spain: Universitat Politècnica de València.

Rowlands, I. (2018). What are we measuring? Refocusing on some fundamentals in the age of desktop bibliometrics. *FEMS Microbiology Letters, 365*(8). doi: 10.1093/femsle/fny059.

Sakaki, T., Okazaki, M., & Matsuo, Y. (2010). *Earthquake shakes Twitter users: real-time event detection by social sensors.* Paper presented at the Proceedings of the 19th international conference on World wide web.

Shaman, J., Karspeck, A., Yang, W., Tamerius, J., & Lipsitch, M. (2013). Real-time influenza forecasts during the 2012–2013 season. *Nature Communications, 4*(1), 2837. doi: 10.1038/ncomms3837.

Signorini, A., Polgreen, P. M., & Segre, A. M. (2010). *Using Twitter to estimate H1N1 influenza activity.* Paper presented at the 9th Annual Conference of the International Society for Disease Surveillance.

Sinnenberg, L., Buttenheim, A. M., Padrez, K., Mancheno, C., Ungar, L., & Merchant, R. M. (2017). Twitter as a tool for health research: A systematic review. *American Journal of Public Health, 107*(1), e1-e8. doi: 10.2105/AJPH.2016.303512.

StataCorp. (2017). *Stata statistical software: release 15*. College Station, TX, USA: Stata Corporation.

Sugimoto, C. R., Work, S., Larivière, V., & Haustein, S. (2017). Scholarly use of social media and altmetrics: A review of the literature. *Journal of the Association for Information Science and Technology, 68*(9), 2037-2062.

Tahamtan, I., & Bornmann, L. (2020). Altmetrics and societal impact measurements: Match or mismatch? A literature review. *El profesional de la información, 29*(1), e290102.

Thelwall, M. (2017). *Web indicators for research evaluation: A practical guide*. London, UK: Morgan & Claypool.

Thelwall, M., & Kousha, K. (2015). Web indicators for research evaluation. Part 2: Social media metrics. *Profesional De La Informacion, 24*(5), 607-620. doi: 10.3145/epi.2015.sep.09.

Triguero, F., Fidalgo-Merino, R., Barros, B., & Fernández-Zubieta, A. (2018). Scientific knowledge percolation process and social impact: A case study on the biotechnology and microbiology perceptions on Twitter. *Science and Public Policy, 45*(6), 804-814. doi: 10.1093/scipol/scy022 %J Science and Public Policy.

Tunger, D., Clermont, M., & Meier, A. (2018). Altmetrics: State of the art and a look into the future. *IntechOpen*. doi: 10.5772/intechopen.76874.

Vainio, J., & Holmberg, K. (2017). Highly tweeted science articles: who tweets them? An analysis of Twitter user profile descriptions. *Scientometrics, 112*(1), 345–366. doi: 10.1007/s11192-017-2368-0.

Vosoughi, S., Roy, D., & Aral, S. (2018). The spread of true and false news online. *Science, 359*(6380), 1146-1151. doi: 10.1126/science.aap9559.

Waltman, L., & Costas, R. (2014). F1000 recommendations as a potential new data source for research evaluation: A comparison with citations. *Journal of the Association for Information Science and Technology, 65*(3), 433-445. doi: 10.1002/asi.23040.

Work, S., Haustein, S., Bowman, T. D., & Larivière, V. (2015). *Social media in scholarly communication. A review of the literature and empirical analysis of Twitter use by SSHRC doctoral award recipients*. Montreal, Canada: Canada Research Chair on the Transformations of Scholarly Communication, University of Montreal.

Wouters, P., Thelwall, M., Kousha, K., Waltman, L., de Rijcke, S., Rushforth, A., & Franssen, T. (2015). *The metric tide: Literature review (supplementary report I to the independent review of the role of metrics in research assessment and management)*. London, UK: Higher Education Funding Council for England (HEFCE).

Wouters, P., Zahedi, Z., & Costas, R. (2019). Social media metrics for new research evaluation. In W. Glänzel, H. F. Moed, U. Schmoch & M. Thelwall (Eds.), *Springer Handbook of Science and Technology Indicators* (pp. 687-713). Cham: Springer International Publishing.

Yu, H. (2017). Context of altmetrics data matters: An investigation of count type and user category. *Scientometrics, 111*(1), 267-283. doi: 10.1007/s11192-017-2251-z.

Zahedi, Z., Costas, R., & Wouters, P. (2014). How well developed are altmetrics? A cross-disciplinary analysis of the presence of 'alternative metrics' in scientific publications. *Scientometrics, 101*(2), 1491-1513. doi: 10.1007/s11192-014-1264-0.

Zubiaga, A., Spina, D., Martínez, R., & Fresno, V. (2014). Real-time classification of twitter trends. *Journal of the Association for Information Science and Technology, 66*(3), 462-473. doi: 10.1002/asi.23186.